\documentclass[aps,twocolumn,preprintnumbers,amsmath,prl]{revtex4}

\usepackage{amsmath} 
\usepackage{amsfonts}
\usepackage{amssymb}
\usepackage{mathrsfs}
\usepackage{rotating}
\usepackage{color}
\usepackage{graphicx}
\usepackage{nicefrac}
\usepackage{framed}

\definecolor{white}{rgb}{1,1,1}
\definecolor{black}{rgb}{0,0,0}
\definecolor{blue}{rgb}{0.1,0.1,1}
\definecolor{cyan}{rgb}{0.5,0.7,0.75}  
\definecolor{red}{rgb}{1,0.1,0.1}
\definecolor{green}{rgb}{0,0.8,0}
\definecolor{purple}{rgb}{0.4,0,0.6}

\newcommand{\blue}{\color{black}}
\newcommand{\cyan}{\color{black}}
\newcommand{\red}{\color{black}}

\newcommand{\purple}{\color{purple}}

\begin{document}


\title{The Vector Space of Convex Curves: How to Mix Shapes} 
%
%

\author{Dongsung Huh} 


\affiliation{Gatsby Computational Neuroscience Unit, University College London, London, W1T 4JG, United Kingdom}

\begin{abstract}
%
We present a novel, log-radius profile representation for convex curves
and define a new operation for combining the shape features of curves. 
Unlike the standard, angle profile-based methods, 
this operation 
accurately combines the shape features in a visually intuitive manner. 
This method have implications in shape analysis 
as well as in investigating how the brain perceives and generates curved shapes and motions. 
%
%
%
\end{abstract}


\maketitle

\section{Introduction}


Understanding how we perceive and understand shape is a central problem in human and computer vision. 
Plane curves are the simplest forms that have shape: we can easily recognize objects in cartoons or outlines of images. 
Here, we introduce a novel way to represent and combine 
the shape of plane curves {\red that has the twin virtues of mathematical elegance and intuitive simplicity.} 
%
%




\section{Angle profile representation} 

A plane curve is formally described as a continuous mapping 
from a closed interval of real numbers to a 2-D Euclidean plane, 
$\Gamma:  \mathbb{I} \to \mathbb{R}^2$, 
{\it i.e.,}  $\Gamma(s)= (x(s),y(s)),$ 
where 
$s \in \mathbb{I} $ is the 1-D coordinate parameterization along the curve.
If the natural, {\it arc-length} coordinate is used 
(such that  $ds = \sqrt{ dx^2 + dy^2}$), 
the derivative of the curve {\blue with respect to $s$} yields a unit-length velocity vector: 
$\Vert {\Gamma}'(s) \Vert =1$. 
%
%
This implies that the curve shape {\blue can be fully characterized} by the velocity vector's  orientation along the curve, {\it i.e.}, the {\it angle profile} $\{\theta(s)\}$,
since the length of the vector does not contain any shape information. 
%



The angle profile representation is widely used for 
%
identification and categorization of curve shapes,
{\blue as well as shape synthesis}
{\purple \cite{zhang2004review}}.
%
%
In particular, 
one of the most widely used shape description methods, called {\it Fourier Descriptors},
analyzes angle profiles of curves
as linear combinations of their frequency components
%
{\purple \cite{cosgriff1960identification, zahn1972fourier}}. 
%
%
{\red However, such 
analysis  
has serious drawbacks,
%
%
because the operation for combining angle profiles 
does not properly translate to combination of shape features.
(See  Appendix~A).}

Here, we introduce a novel, {\it log-radius profile} representation 
that is dual to the angle profile representation.
This representation resolves the problems of the angle profile representation,
and provides a new operation for combining shape features. 

\begin{figure}[b]
\begin{center}	
\includegraphics[width=0.49\textwidth]{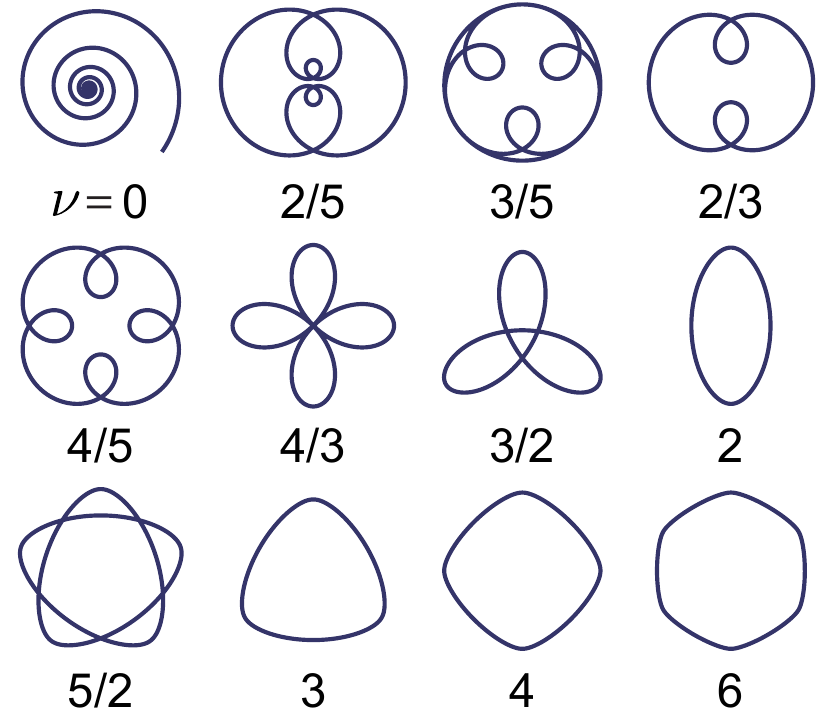}
\end{center} 
\caption[]{Examples of elementary shapes shown with their characteristic frequency $\nu$. 
See {\purple  https://www.youtube.com/watch?v=waXWOv0YqFE} 
for a movie showing how the shape varies with continuously changing frequency.
}
\label{fig:Elementary_shapes}
\end{figure}

\section{Log-Radius profile representation of Convex curves}


%
%
Consider a subset of plane curves with monotonically increasing angle profiles, 
called {\it convex curves}.
%
%
%
Such a curve admits 
an alternative, angle-based parameterization,  $\Gamma(\theta)$,
since any point on the curve can be uniquely specified by $\theta$. 
%
The angle coordinate, like arc-length, is a natural parameterization of a curve,
determined uniquely by the curve's geometry.  
Moreover, it is scale invariant 
| the coordinate value of a point 
remains unchanged under scaling operations on the curve.
%

Differentiating a curve {\blue with respect to $\theta$} 
yields a velocity vector 
whose length is the local radius of curvature: 
	\begin{equation} 
		\Vert {\Gamma}'(\theta) \Vert =  \frac{ds}{d\theta} \equiv r(\theta)  ~ {\cyan >0}.
	\end{equation}
Then, 
the shape of a convex curve {\blue can be fully characterized} by the {\it radius profile} $\{r(\theta)\}$ along the curve, because the orientation of the velocity vector only provides redundant information that is already specified by the angle coordinate. 
%

To summarize, the radius-profile describes the length of the velocity vector as a function of the orientation, 
whereas the angle-profile describes the orientation of the velocity vector as a function of the arc-length: They provide complementary ways to represent curves. 

However, since we will consider scaling operations on curves, it is more {\red convenient} to introduce the {\it log-radius profile} representation, 
$\{l(\theta)\}$, defined as
	\begin{equation} \label{eq:log-radius}
		l(\theta) \equiv \log r(\theta).
	\end{equation}


\section{Elementary shapes} 



A circle is a simple featureless curve, 
described by a constant log-radius of curvature. 
{\red However, interesting shape features are described by fluctuations of the log-radius profiles.}

Let us define {\it elementary shapes} 
by sinusoidal log-radius profiles:
%
%
%
\begin{align} \label{eq:elementary_shapes}
	l(\theta)  = \epsilon \sin(\nu(\theta-\theta_o)),  
\end{align}
%
where $\nu$ is the frequency, $\epsilon$ is the amplitude 
and $\theta_o$ is the phase shift, which rotates the shape
(Fig.~\ref{fig:Elementary_shapes}).
%

Each elementary shape exhibits a distinctive feature characterized by the frequency $\nu$. 
%
%
%
For example, {\cyan the elongated shape of} an elliptic curve is characterized by frequency 2, whose log-radius profile oscillates twice per one full rotation of $\theta$, or $2\pi$ radians.  
At larger integer frequencies,  
the shapes resemble rounded regular polygons.
%
In general, an elementary shape with a rational frequency $\nu = m/n$, 
where $m$ and $n$ are coprime integers ({\it i.e.} no common factors) and $m > 1$, 
has a closed shape of period  $\Theta = 2 \pi n$, and exhibits $m$ degrees of rotational symmetry. 
If $m=1$, then the curve {\cyan does not close and} exhibits a translational symmetry.
%
%
In the zero frequency limit, the elementary shape approaches a logarithmic spiral: 
$l(\theta) = \lim_{\nu \to 0} (a /\nu) \sin(\nu \theta) = a\theta$.

\begin{figure}
\begin{center}	
\includegraphics[width=0.49\textwidth]{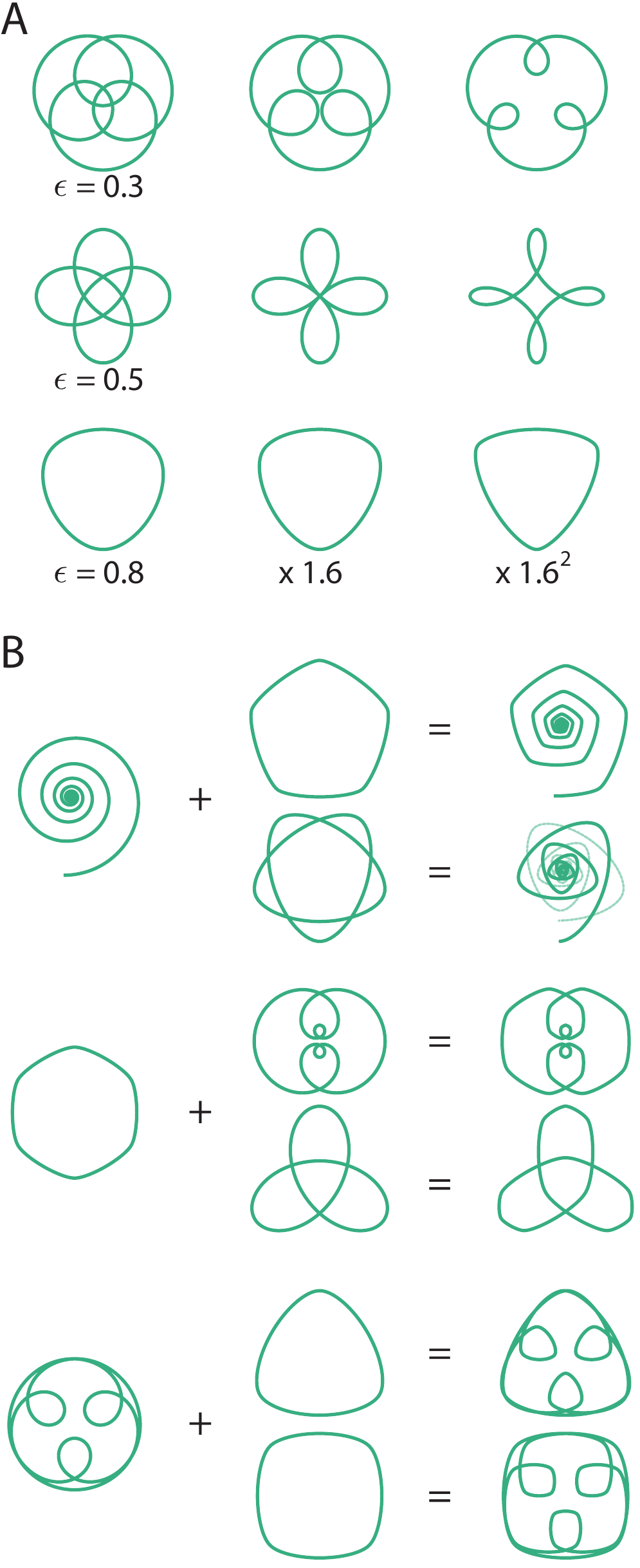}%
\end{center} 
\caption{(A) Scalar multiplication of curves, eq~\eqref{eq:Scalar_multiplication}. Size is normalized.
	(B) Addition operation between two curves, eq~\eqref{eq:Addition}. 
This operation accurately combines the shape features of curves. 
}
\label{fig:Scalar_multiplication_Addition}
\end{figure}

The amplitude $\epsilon$ modulates {\blue the degree of expression} of elementary shapes,
while preserving their characteristic features (see Fig~\ref{fig:Scalar_multiplication_Addition}A). 
%
%

%
%


\section{Vector Space of convex curves} 


Uniform scaling is the simplest operation on a curve,
which scales the overall size of the curve while preserving its shape. 
{\red In our representation}, this corresponds to adding a constant to the log-radius profile.



More generally, we define two operations on convex curves, 
{\red $\Gamma_i:  \mathbb{I} \to \mathbb{R}^2$}:
scalar multiplication 
(Fig~\ref{fig:Scalar_multiplication_Addition}A)
\begin{equation}		\label{eq:Scalar_multiplication}
	\Gamma = a \cdot \Gamma_o  ~ \longleftrightarrow ~  l(\theta) = a \cdot l_o(\theta),  
\end{equation}
which amplifies the log-radius profile by a scalar factor
$(a\in\mathbb{R}, \forall \theta \in \mathbb{I})$,
and addition between two curves 
(Fig~\ref{fig:Scalar_multiplication_Addition}B)
 \begin {equation} 		\label{eq:Addition}
 	 \Gamma = \Gamma_{1} + \Gamma_{2} ~ \longleftrightarrow  ~ l(\theta) = l_1(\theta) +  l_2(\theta).
\end{equation}
which adds their log-radius profiles in a pointwise manner
$(\forall \theta \in \mathbb{I})$.


The addition operation {\blue can be understood as} 
a generalized scaling operation, 
which uses one curve's log-radius profile as the scale factor for modifying the other curve.
%
It reduces to uniform scaling 
{\blue if one of the curves is a circle}.
The unit circle, $ l(\theta) = 0 $, is the identity element of the addition operation. 

Remarkably, the addition operation {\red exquisitely} merges the shape features of the added curves
in a visually intuitive manner 
(See Fig~\ref{fig:Scalar_multiplication_Addition}B).
For examples, adding a shrinking spiral progressively decreases the length scale of the curve, thereby ``spiralizing" its shape, 
whereas adding an ellipse elongates the curve in one direction and compresses it in the perpendicular direction, thereby ``elliptizing" the curve. 
{\red Thus, adding a spiral and an ellipse produces 
an elliptic spiral, which combines the shape features of both curves.} 
This {\blue accurate} combination of shape features {\red owes to} 
the angle coordinate representation, which is invariant under scaling operations.

The {\red scalar multiplication and addition operations} define a vector space over the set of convex curves, which is spanned by the basis set of elementary {\red shape curves} eq~\eqref{eq:elementary_shapes}; 
that is, any curve in this space can be {\red represented as} a linear combination of 
elementary curves. 
Moreover, an inner-product between curves can be defined as,
\begin{equation} \label{eq:innerproduct}
	\langle \Gamma_1, \Gamma_2 \rangle  \equiv 
	 \int_{\theta\in \mathbb{I}} 
						l_1(\theta)  \, l_2(\theta) \, d\theta,
\end{equation}
which induces a norm $\Vert \Gamma \Vert \equiv \sqrt{ \langle \Gamma, \Gamma \rangle}$.
{\red Thus,  convex curves with finite norm ($\Vert \Gamma \Vert < \infty $)
form a Hilbert space, isomorphic to the space of square-integrable functions, $L_2$.}



\section{Discussion}


We presented a novel, log-radius profile representation for describing convex shapes,
which is dual to the standard, angle profile-based representation
%
and offers a complementary view of curve shapes. 
%
%
The angle profile representation is closely related to 
``bending" operations:
the simplest curve is a straight line, 
which can be bent into various curves. 
%
%
In contrast, the log-radius profile representation is closely related to 
``scaling" operations:
the simplest curve is a circle, 
which can be non-uniformly scaled 
into other curve shapes.
%
%

The log-radius profile representation resolves the {\cyan aforementioned} problems of {\cyan the} angle profile representation
(See Appendix~A):  
The elementary curves 
{\blue preserve} their characteristic shape features over all range of amplitude,
and the addition operation 
accurately combines {\cyan the} shape features. 
%
Therefore, Fourier transform of log-radius profiles {\red indeed properly analyzes the curve shapes 
into visually meaningful shape features.} 

%


Recent applications of this method have revealed surprising details of regularities in kinematics of 
curved hand movements {\purple \cite{huh2015spectrum}},
as well as in speed perception of curved motion {\purple \cite{huh2013novel}}.
It may also be useful in investigating the shape perception process in vision. 

Here, we considered representation of 1-D convex curves in 2-D space. This result can be generalized to higher dimensional, convex surfaces. The angle coordinate then becomes related to normal vector to the surface and inverse Gauss map. 

\section*{Acknowledgement}
I thank Terrence J. Sejnowski for helpful comments. 
This research was supported by Gatsby Charitable Foundation.

\section*{Appendix A: Fourier Descriptors}

\makeatletter 
\setcounter{equation}{0} 
\setcounter{figure}{0} 
\renewcommand{\theequation}{A\arabic{equation}}
\renewcommand{\thefigure}{A\@arabic\c@figure}
 \makeatother


The Fourier Descriptor method describes the angle profile of a curve 
as a linear combination of frequency components
%
\begin{equation} \label{eq:Fourier_Descriptors}
	\theta(s) = s \Theta + \sum_{k=1}^{\infty} a_k \cos( k s \Theta) + b_k \sin( k s \Theta),
\end{equation}
where $s \in [0,1]$ is the normalized arc-length coordinate and $\Theta$ determines the total number of turns ($2\pi$ for simple closed curves)
%
{\purple \cite{zahn1972fourier}}.
The coefficients $a_k, b_k$ are called Fourier descriptors.

\begin{figure}[t]
\begin{center}	
\includegraphics[width=0.45\textwidth]{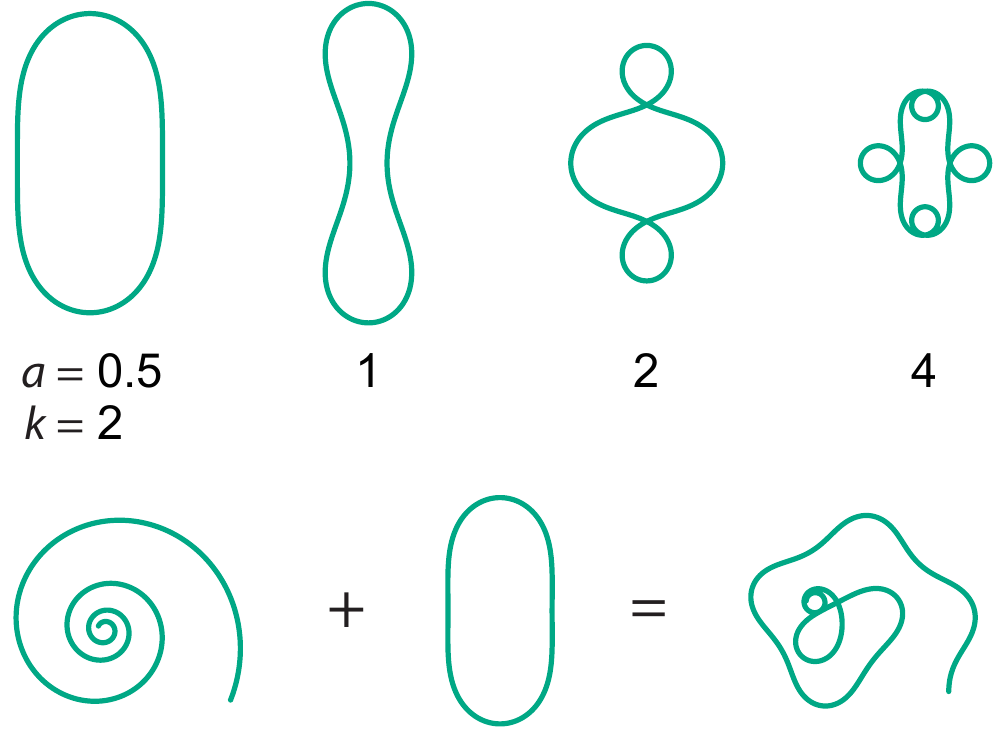}%
\end{center} 
\caption{Problems of angle profile representation. 
Top: A single frequency component shape 
described by eq~\eqref{eq:angle_pure_frequency}. 
Bottom: Addition operation in angle profile representation.}
\label{fig:angle_profile_failure}
\end{figure}

However, such approach has serious drawbacks.
First, 
the curve shape described by a single frequency component 
\begin{equation} \label{eq:angle_pure_frequency}
	\theta_k(s) =s \Theta + a_k \cos( k s \Theta) 
\end{equation}
exhibits a wildly varying shape as the coefficient $a_k$ increases 
(Fig~\ref{fig:angle_profile_failure}, top),
thus failing to represent a unique, consistent shape feature.
Secondly, in the angle profile representation, linear addition of the frequency components 
does not properly combine their shape features, 
but tends to produce 
rather deformed shapes 
(Fig~\ref{fig:angle_profile_failure}, bottom).
%
%
%
Therefore, decomposing an angle profile into the frequency components via Fourier Transform 
does not {\cyan properly} translate to a meaningful analysis of the curve's shape into basic shape features. 


%
%

%


\end{document}